\documentstyle[prd,aps,floats]{revtex}
\begin{document}
\draft

%
%
\input epsf \renewcommand{\topfraction}{0.8}
\twocolumn[\hsize\textwidth\columnwidth\hsize\csname
@twocolumnfalse\endcsname

\title{Preheating in Generalized Einstein Theories} \author{Anupam
  Mazumdar and Lu\'{\i}s E. Mendes} \address{Astrophysics Group,
  Blackett Laboratory, Imperial College London, SW7 2BZ, ~~~U.~K.}
\date{\today} \maketitle
\begin{abstract}
  We study the preheating scenario in Generalized Einstein Theories,
  considering a class of such theories which are conformally
  equivalent to those of an extra field with a modified potential in
  the Einstein frame.  Resonant creation of bosons from oscillating
  inflaton has been studied before in the context of general
  relativity taking also into account the effect of metric
  perturbations in linearized gravity. As a natural generalization we
  include the dilatonic/Brans-Dicke field without any potential of its
  own and in particular we study the linear theory of perturbations
  including the metric perturbations in the longitudinal gauge. We
  show that there is an amplification of the perturbations in the
  dilaton/Brans-Dicke field on super-horizon scales ($k \rightarrow 0
  $) due to the fluctuations in metric, thus leading to an oscillating
  Newton's constant with very high frequency within the horizon and
  with growing amplitude outside the horizon. We briefly mention the
  entropy perturbations generated by such fluctuations and also the 
  possibility to excite the Kaluza-Klein modes
  in the theories where the dilatonic/Brans-Dicke field is interpreted
  as a homogeneous field appearing due to the dimensional reduction
  from higher dimensional theories.
\end{abstract}

\pacs{PACS numbers: 98.80.Cq \hspace*{1.3cm} Imperial preprint
  Imperial-AST 99/2-1, hep-ph/yymmmnn}

\vskip2pc]


\section{Introduction}

One of the most important epochs in the history of inflationary
Universe is the transition from almost De Sitter expansion to the
radiation dominated universe. Until few years ago proper understanding
of this phenomena was not well established. Recent development in the
theory of resonant particle creation of bosons~\cite{linde} as well as
fermions~\cite{greene} due to coherent oscillations of the
Bose-condensate inflatons may explain satisfactorily the emergence of
radiation era from the ultra cold inflationary universe~\cite{linde}.
Although such phenomena are hard to reproduce in a laboratory they are
the simplest manifestation of a slowly varying scalar field which
rolls down the potential and oscillates as a coherent source at each
and every space time region.  Apart from reheating the universe to the
temperature ambient for the production of light nuclei this phenomena
has multifaceted consequences.  It is also a well known mechanism to
create non-equilibrium environment which can be exploited for the
generation of net baryon antibaryon asymmetry required for the
baryogenesis~\cite{kolb}. If there exist general chiral fields, then,
in particular, breaking of parity invariance could also lead to
different production of left and right fermions.  If the rotational
invariance is broken explicitly by an axial background then there
would be anisotropic distribution of fermions~\cite{anupam}.  The same
phenomena is also responsible for the generation of primordial
magnetic field~\cite{giov,gi} in the context of string cosmology,
where the dynamical dilaton field plays the role of an oscillating
background.

The preheating scenario has been studied so far in the context of
general theory of relativity. It has been well established from the
present observations that this theory of gravitation is the correct
description of space-time geometry at scales larger than $1$ cm. There
exists a class of deviant theories which are scalar-tensor gravity
theories, known as Generalized Einstein Theories (GET) of which the
Jordan--Brans--Dicke (JBD) theory~\cite{JBD,Will} is the simplest and
best-studied generalization of general relativity. This theory leads
to variations in the Newtonian gravitation "constant" $G$, and
introduces a new coupling constant $\omega$, with general relativity
recovered in the limit $1/\omega \rightarrow 0$.  The constraint on
$\omega$ based on timing experiments using the Viking space probe
suggests that it must exceed $500$~\cite{reas}. The JBD theory also
mimics the effective lagrangian derived from low energy scale of the
string theory where the Brans-Dicke (BD) field is called Dilaton and
the coupling constant $\omega $ takes the negative value of 
$-1$~\cite{string}. It also represents the $(4+D)$ dimensional
Kaluza--Klein theories with an inflaton field which has mainly two
subclasses out of which we shall consider the one where the inflaton
is introduced in an effective four dimensional theory. In this case the
BD/dilaton field plays the role of homogeneous scalar field in $4$
dimensions and is related to the size of the compactification. Most of
these models are conformally equivalent and can be recast in the form
of Einstein gravity theory~\cite{conf}. The only difference is that
the scaling of the fields and their corresponding couplings will be
different for different interpretations of BD/dilaton field.

In this paper we shall consider the general relativity limit of the
JBD theory as well as other variant theories such as string and
Kaluza--Klein and for the sake of consistency we just use one
representative of the coupling constant, $\gamma $, which takes
different values accordingly. We must say that there is an obvious
advantage in all these theories that they are well constrained at the
present day and the evolution of the BD/dilaton field is roughly
constant after a period of $60$ e-foldings of inflation. Here we
shall not consider the potential in the dilatonic/BD sector and we
assume that the evolution of the dilaton/BD field during the
oscillatory phase is solely due to its coupling with the inflaton in
the Einstein frame. We assume the quadratic potential, $V(\phi) =
\frac{1}{2}m^2 \phi ^{2}$ for the inflaton field. It is to be noticed
that during the oscillatory phase the field $\phi$ decreases in the
same way as the density of non-relativistic particles of mass $m :
\rho_{\phi} = \frac{1}{2}\dot \phi^2 + \frac{1}{2}m^2 \phi^2 \approx
a^{-3}$, provided the coupling $\gamma $ is very small as in the case
of JBD. For simplicity we discuss the physics of weak coupling and at
the end we comment on the strong coupling limits which we get
numerically. In the weak limit the coherent oscillations of the
homogeneous scalar field correspond to a matter dominated equation of
state with vanishing pressure. This suggests that BD field should
evolve according to a well known solution during the dust
era~\cite{weinberg}.

Apart from the study of non-perturbative creation of bosons and
fermions during preheating, metric perturbations have also been studied
extensively in~\cite{bassett1} and~\cite{fin}. It has been shown that
the metric perturbations also grow during preheating in the
multi-scalar case due to the enhancement of the entropy perturbations.
It is also possible to amplify the super-horizon modes causally. Such
amplification requires the linear theory of gravitational
perturbations to be supplanted by non-linear perturbation theory. 
In~\cite{bassett1} the authors have discussed the two fields case in
particular where the first one ($\phi$) represents the inflaton and
the second ($\sigma$) represents the newly created bosons with an
interaction with the inflaton of the form $\frac{1}{2}g \phi ^2
\sigma ^2$. Perturbations in $\sigma $ get amplified along with the
metric perturbations. They have argued that all the fields would be
possibly amplified except the inflaton which is subdued by
transferring its energy to the other fields. Following their claim it
is also possible to amplify the other fields such as dilaton/BD field
as they are also coupled to the inflaton and can be cause of some concern.
In the concluding section we devote ourselves to the discussion of
these issues.

\section{The equations}

In JBD theory the action in the Jordan frame~\cite{JBD}
\begin{eqnarray}
  \label{jbdeq}
  S=\int \left[\frac{\Phi_{BD}}{16\pi}\hat R + \frac{\omega ^2}{16 \pi
      \Phi_{BD}} 
    \hat g^{\mu \nu} \partial _\mu \Phi_{BD} \partial_\nu \Phi_{BD}\,
  \right. \nonumber \\
  \left. +\frac{1}{2}\hat g^{\mu \nu} \partial_\mu \phi \partial_\nu \phi -
    V(\phi) \right] \sqrt{-\hat g} d^4 x \,,
\end{eqnarray}
is transformed into Einstein frame
\begin{eqnarray}
\label{lag}
S=\int\left[\frac{1}{2 \kappa^2}R +\frac{1}{2} g^{\mu \nu} \partial _\mu
\chi \partial_\nu \chi -U(\chi) + \,
\right. \nonumber \\
\left.  \frac{1}{2} e^{-\gamma \kappa \chi}
 g^{\mu \nu} \partial _\mu \phi \partial_\nu \phi -
e^{-2\gamma \kappa \chi}V(\phi) \right]\sqrt{-g}d^4 x \,,
\end{eqnarray}
through the conformal transformation
\begin{eqnarray}
\label{con}
g_{\mu \nu}&=& \Omega^2 \hat g_{\mu \nu} \,, \nonumber \\
\Omega^2 &\equiv & \frac{\kappa^2}{8 \pi} \Phi_{BD} \equiv
e^{(\kappa \chi / \sqrt{\omega +3/2})} \,,
\end{eqnarray}
where $\kappa ^2 =8\pi G$, $\gamma$ is a constant related to $\omega
$ and $\chi $ and $\phi $ are the BD and inflaton fields respectively.
For our concern $\gamma = \frac{1}{\sqrt{\omega + 3/2}}$ and $U(\chi)
=0$ (see however the discussion below and on Sec.~\ref{sec:conc}). 
Observational constraints give $\omega > 500$ so that $\gamma <0.09 
\ll 1$. In the dimensionally reduced theories the original action is
different from Eq.~(\ref{jbdeq}) but it is still possible to 
conformally transform it to Eq.~(\ref{lag})~\cite{conf}. In this case
$\chi$ and the coupling $\gamma$ are given by 
\begin{eqnarray}
\label{kal}
\chi &=& \left[\frac{2(D +2)}{D}\right]^{1/2} \ln \Phi_{BD} \,,\\
\gamma &=& \left[\frac{2D}{D+2}\right]^{1/2} \,,
\end{eqnarray}
where $\Phi_{BD} \equiv (b/b_0)^{D/2}$, $b$ is the radius of
compactification and $b_0$ its present value. Notice that in the
Kaluza-Klein case it is possible to make $U(\chi)=0$ only when the
extra dimensions are compactified on a torus which has zero
curvature. Compactifying the extra dimensions on a sphere will give
rise to a mass to the dilaton corresponding to the curvature of the
sphere. In Sec.~\ref{sec:conc} we 
shall come back to this point again and discuss what should we expect 
if such a term is also included in the lagrangian.  The number of extra
dimensions is $D$ and $\gamma $ can at most take a value $\sqrt{2}$.
In the superstring case $\gamma $ is exactly $\sqrt{2}$~\cite{berkin}. As
we have already mentioned during radiation and matter domination
$\chi$ has to be roughly constant in order to
comply with the observational data~\cite{barrow} and to reproduce the
correct value of the gravitational constant today.  In most of the
models the evolution of the extra dimensions is also treated to be
constant during the radiation era~\cite{gasp}.

\subsection{Equations of Motion}

We shall perform our calculation in the Einstein frame,
Eq.~(\ref{lag}), for simplicity. The homogeneous equations of motion
for a zero--curvature Friedmann universe are
\begin{eqnarray}
\label{heq1}
\ddot \chi + 3H\dot \chi + \frac{\gamma}{2}e^{-\gamma \chi}\dot \phi^2
-2\gamma e^{-2\gamma \chi} V(\phi)&=& 0 \,, \\
\label{heq2}
\ddot \phi +3H\dot \phi - \gamma \dot \chi \dot \phi +
e^{-\gamma \chi}  V^{\prime}(\phi) &=&0 \,, \\
\label{heq3}
\frac{1}{3}\left[\frac{1}{2}\dot \chi^2 + \frac{1}{2}e^{-\gamma \chi}
\dot \phi^2 + e^{-2\gamma \chi} V(\phi) \right] &=& H^2\, ,
\end{eqnarray} 
with an overdot denoting time derivation and a prime denoting partial
derivative with respect to $\phi $. We use natural units where
$\kappa^2 =1$.  We do not pay much attention to the field equations
during inflation, since much has been studied in this 
area~\cite{berkin,starobinsky}, rather we concentrate on the coherent
oscillations of the inflaton.  Before we turn to linear perturbation
theory we should mention that we have to introduce the bosons
resonantly produced from the non-perturbative decay of inflaton field
which we shall denote by $\sigma$. We consider the coupling of
$\sigma$ particles with the inflaton to be $\frac{g}{2}\phi ^2 \sigma
^2$, where $g$ is the coupling strength.  Hence, the old potential is
modified during oscillations due to the presence of massless $\sigma $
particles:
\begin{equation}
V(\phi , \sigma ) = \frac{1}{2}m^2 \phi ^2 + \frac{1}{2}g\phi ^2 \sigma ^2 \,.
\end{equation}
In the Heisenberg representation the equation of motion for the scalar
field $\sigma $ can be expressed in terms of the temporal part of the
mode with comoving momentum $k$,
\begin{equation}
\label{hom}
\ddot \sigma +3H \dot \sigma -\gamma\dot \chi \dot \sigma
+\left [ \frac{k^2}{a^2(t)}+ g\phi^2 e^{-\gamma \chi} 
-\zeta R\right ]\sigma  =0 \, ,
\end{equation}
where $\zeta $ is the nonminimal coupling to the curvature, which we
shall not take into account. This is the equation of great importance
for the study of resonant creation of $\sigma $ (details can be seen
in~\cite{linde}). This particular equation takes the form of the well known
Mathieu equation after redefining the field. This equation has very
rich properties, and its solutions can fall into two categories,
either stable or unstable depending on the choice of two parameters
$(A(k),q)$, defined later~\cite{mathe}.

\subsection{ Linear Perturbations}

Linear perturbations can be taken into account in a gauge invariant
fashion using the longitudinal gauge. We write the perturbed metric as
\begin{eqnarray}
ds^2 = (1+2\Phi )dt^2 -a^2(t)(1-2\Psi)\delta _{ij}dx^{i}dx^{j}\,,
\end{eqnarray}
where $\Phi$ and $\Psi$ are gauge invariant metric potentials. The
Fourier modes satisfy the following equations of motion which are
derived perturbing the Einstein's equations~\cite{bardeen,mukh}.
\begin{eqnarray}
\label{pert1}
\Phi = \Psi \,, 
\end{eqnarray}
\begin{eqnarray}
\label{pert2}
\dot \Phi + H\Phi = \frac{1}{2}\left[\dot \chi \delta \chi + 
e^{-\gamma \chi}\dot \phi \delta \phi + e^{-\gamma \chi}\dot \sigma \delta 
\sigma \, \right] \,, 
\end{eqnarray}
\begin{eqnarray}
\label{pert3}
\delta \ddot \chi +3H \dot \delta\chi
+\left[\frac{k^2}{a^2}-\frac{\gamma ^2}{2} 
e^{-\gamma \chi}\dot \phi^2 
+4\gamma ^2 e^{-2\gamma \chi }\cdot 
V(\phi ,\sigma )\right] \delta \chi  
\nonumber  \\ 
+\gamma e^{-\gamma \chi}\dot \phi \delta \dot \phi 
-2\gamma e^{-2\gamma \chi }V(\phi ,\sigma )_{,\phi }\delta \phi  
-2\gamma e^{-2\gamma \chi }V(\phi ,\sigma )_{,\sigma }\delta \sigma = 
\nonumber \\
4\dot \Phi \dot \chi +4e^{-2\gamma
\chi}V(\phi ,\sigma )\Phi \,,  
\end{eqnarray}
\begin{eqnarray}
\label{pert4}
\delta \ddot \phi +(3H-\gamma \dot \chi )\delta \dot \phi +\left[
\frac{k^2}{a^2}+
e^{-\gamma \chi }V(\phi ,\sigma )_{,\phi ,\phi }\right]\delta \phi         
\nonumber  \\
+e^{-\gamma \chi }V(\phi ,\sigma )_{,\phi ,\sigma}\delta \sigma
-\gamma \dot \phi \delta \dot \chi -\gamma e^{-\gamma \chi}
V(\phi ,\sigma )_{,\phi }\delta \chi =
\nonumber  \\
4\dot \Phi \dot \phi -2e^{-\gamma \chi }V(\phi ,\sigma )_{,\phi }\Phi \,,
\end{eqnarray}
\begin{eqnarray}
\label{pert5}
\delta \ddot \sigma +(3H-\gamma \dot \chi )\delta \dot \sigma +
\left[\frac{k^2}{a^2}+g\phi ^2 e^{-\gamma \chi }\right]\delta \sigma 
\nonumber \\
+2g\sigma \phi e^{-\gamma \chi }\delta \phi -\gamma \dot \sigma \delta \dot
\chi -\gamma g\phi ^2 \sigma e^{-\gamma \chi}\delta \chi =
\nonumber \\
4\dot \Phi \dot \sigma -2g\phi ^2 \sigma e^{-\gamma \chi }\Phi \, ,
\end{eqnarray}
where $\delta \phi, \delta \chi $ and $\delta \sigma $ are gauge
invariant perturbations in the respective fields. At this point the
evolution equations for the background are not determined by
Eqs.~(\ref{heq1}--\ref{heq3}); rather they are modified by the
presence of $\sigma$ particles. For the Friedmann equation
(Eq.~(\ref{heq3})) we now have
\begin{eqnarray}
\frac{1}{3}\left[\frac{1}{2}\dot \chi^2 + \frac{1}{2}e^{-\gamma \chi}
\dot \phi^2 +\frac{1}{2}e^{-\gamma \chi }\dot \sigma ^2 + 
e^{-2\gamma \chi} V(\phi, \sigma ) \right] &=& H^2\,.
\end{eqnarray} 
Similarly, the homogeneous equations for $\chi $ and $\phi $,
Eqs.~(\ref{heq1}) and (\ref{heq2}) are to be modified accordingly
with $V(\phi )$ replaced by
$V(\phi, \sigma )$ and $V^{\prime}(\phi )$ replaced by $V(\phi
,\sigma )_{,\phi }$.  It is worth mentioning that in the absence of
$\sigma $ field and the metric perturbation it is possible to express
the perturbation in the BD field in a simpler form. Let us further
note that in an expanding universe without BD field the behaviour of
the inflaton field is $\phi \approx \frac{{\sin }(m_{\phi
    }t)}{m_{\phi}t}$ and the scale factor goes like $a(t)\approx
a_{0}(\frac{t}{t_0})^{2/3}$.  In the presence of the BD field this is 
modified to $a(t)\approx a_{0} (\frac{t}{t_0})^{(2\omega +3)/(3\omega
  +4)}$, such that in the large $\omega $ limit it reduces to the
previous one.  We expect similar modification in the behaviour of an
inflaton field. For our calculation sake we can assume that the
modified evolution of the inflaton for $\omega > 500$ and $\gamma <
0.09 $ is
\begin{eqnarray}
\label{sol}
\phi \approx (1+{\cal O}(\gamma ))\frac{\sin(m_{\phi }t)}{m_{\phi }t} \,.
\end{eqnarray}
For our purpose the exact numerical expression for ${\cal O}(\gamma )$
does not matter because we are retaining the lowest order in $\gamma
$. Simplifying the perturbed BD field Eq.~(\ref{pert3}) with the
help of Eq.~(\ref{sol}) and noting that $\delta \chi $ can be rescaled
by introducing $u =a^{3/2}(t) \delta \chi (t)$ and $z=m_{\phi }t $, we
get
\begin{equation}
\label{master}
\frac{d^2 u}{dz^2} + \left[ \frac{k^2}{a^2 m_{\phi}^2} +
\frac{3}{4} \frac{\gamma ^2}{z^2}
-\frac{5}{4} \frac{\gamma ^2}{z^2} \cos(2z)\right ]u=0 \,.
\end{equation}
The above equation resembles the Mathieu equation~\cite{mathe}:
\begin{equation}
x^{\prime \prime }+\left[A(k) -2q \cos2z \right]x =0\,.
\end{equation}
with
\begin{eqnarray}
A(k)&=& \frac{k^2}{a^2 m_{\phi }^2}+\frac{3}{4}\frac{\gamma ^2}{z^2}\,,\\
q &=& \frac{5}{8}\frac{\gamma ^2}{z^2} \,.
\end{eqnarray}
Since $q \ll 1$ we are never in the broad resonance regime which is
essential to amplify the modes during the oscillations of the scalar
field in presence of expansion. In this case we are always in the
narrow resonance regime~\cite{linde}.

During preheating the resonant modes should grow as $\delta \chi_{k}
\propto exp(\mu _{k}t)$, where $\mu_{k}$ is the Floquet index and its
analytical estimation has been done in~\cite{linde} for the bosonic
fields, $\mu_k \approx \ln(1+2e^{-\pi (k/ag\phi_0)^2}-Q)$, where $Q$
is contributed by the initial and final quantum states of the created
bosons, which in the case of preheating are not thermal but 
have the characteristics of a squeezed state~\cite{linde}. As it has
been pointed out in~\cite{bassett1}, the resonant production of newly
created bosons grows for $k=0$, so the super-Hubble modes are present
even in the absence of metric perturbation. In this case however, the
BD modes are not amplified. We recognize that in the Einstein
equations as we perturb the matter sector we automatically perturb the
gravitational sector, and gravity being a source of negative heat
capacity it can only transfer energy to the metric perturbation which
acts back to the matter sector enhancing the resulting perturbation in
the matter fields. Energetically it is easier to transfer energy to
the lowest modes so most of the energy is transfered to $\Phi_{k=0}$
and that is the reason why $k=0$ is favoured.
\subsection{Numerical Result}
Our analytical approximation breaks down as we increase the numerical
value for the coupling constant $\gamma $ and the perturbation
equations become intractable as we introduce the metric perturbations.
Hence we solve the system of fourteen first order differential
equations numerically. Here while solving the homogeneous equation for
$\sigma $ we consider only the zero mode contribution, hence
Eq.~(\ref{hom}) can be treated as a homogeneous background equation for
the linear perturbation in $\sigma$ with $k=0$. By doing so we neglect
the zero mode contribution from the metric potential and including
the zero mode metric contribution can only lead to enhancement in the 
amplification in the matter and the BD/dilaton sector as discussed in 
\cite{bassett1}. For our 
numerical calculation we have assumed that the perturbation in $\phi$ 
and in $\chi$ contains the generic,
scale invariant spectrum produced during inflation. We set
$\Phi_{k}(t_0) =10^{-5}$ and we also consider the same initial
conditions for the fluctuations in the BD/dilaton and the fluctuations
in the matter fields $\delta \chi_{k=0}=\delta
\phi_{k=0}=\delta\sigma_{k=0}=10^{-5}$ while the  corresponding time
derivatives are set to zero\footnote{We have chosen
  our initial conditions to be scale-invariant i.e. independent of
  k. There are however other subtleties involved in deciding
  the initial conditions. For a recent discussion we
  refer to~\cite{bassett1}.}. The initial condition for $\sigma $ is that of a
plane wave solution and the BD/dilaton field $\chi $ is assumed to be
very small $\ll 1$ to match the present strength of the gravitational
constant.

We have plotted the fluctuations in the BD/dilaton field
and the perturbations in the $\sigma $ field for $k=0$. In the weak
limit for $\gamma =0.09$, the perturbations do not grow much and they
saturate much earlier in comparison with the strong coupling limit, such as
$\gamma=1.22$, which, in the case of a Kaluza-Klein theory corresponds
to $6$ extra dimensions compactified to
a six dimensional torus. In the strong limit case the non-linearity
is achieved as soon as the metric perturbation $\Phi =1 $. It is
obvious that for Kaluza-Klein theories and for string motivated
theories the linear theory soon becomes invalid unless the
backreaction is taken into account. It is important to notice that
our $q$ parameter is not the same as in general relativity. Instead of
defining the $q$ parameter to be $ q= g\frac{\phi_0 ^2}{m_{\phi }^2}$,
where $\phi_0 $ is the initial amplitude of the inflaton field, which
is $0.08 M_{Pl}$ in our case, in presence of the BD field $q$ is modified
by an exponential factor $q=g\frac{\phi_0 ^2}{m_{\phi }^2} e^{-\gamma
  \chi }$.  The negative sign in the exponent drags down the $q$
parameter. For the above figures $g\frac{\phi_0 ^2}{m_{\phi }^2 } =
9.10^3$.  There is some difference between the perturbations in
$\sigma $ and $\chi $ as visible from the plots.  It is important to
note that $\chi $ has as such no potential with minima unlike $\phi $
and $\sigma $ fields which oscillate around the minima of their
respective potentials and for the modes outside the horizon the
perturbed $\chi $ field almost stops oscillating and starts growing
exponentially. The increment in the perturbations in the strong
coupling limit can be understood qualitatively. Let us first see how
the scale factor behaves in these two cases. As we have already
discussed the oscillations in the inflaton field on average correspond
to zero pressure.  This suggests that the BD/dilaton field on average
evolves similar to the matter dominated era, and in presence of the 
BD/dilaton $ a(t) \approx a_0(\frac{t}{t_0})^ {(2\omega +3)/ (3\omega
  +4)}$. For large $\omega $ and small $\gamma $, the growth in $a(t)$
is roughly the same as in general relativity, but for small
$\omega $ and $\gamma $ close to $\sqrt{2}$, the departure of the
behaviour of the inflaton oscillations from general relativity is quite
significant. Since $\phi\sim 1/t \sim 1/a^{(3\omega +4)/ (2\omega
  +3)}$, for small $\omega$, the amplitude of the inflaton 
decreases slowly compared to the general relativity limit, therefore
leading to a remarkable growth in the production of $\sigma $ particles.
\begin{figure}[t]
  \centering
  \leavevmode\epsfysize=6cm \epsfbox{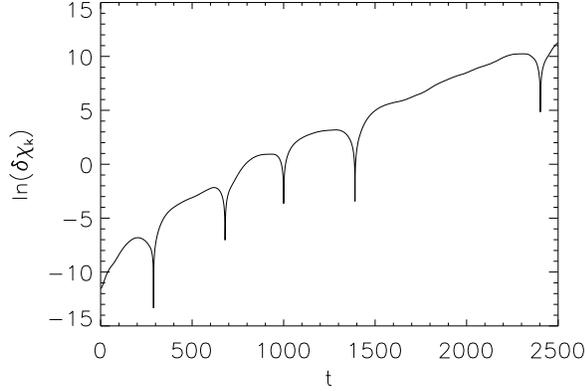}\\
\caption[fig1]{ The evolution of $\ln \delta \chi $ for 
  $g\frac{\phi _0 ^2}{m_{\phi }^2} =9.10^3$, $k=0$ and $\gamma =1.22$
  corresponds to $D=6$ extra compactified dimensions.  }
\end{figure}
\begin{figure}[t]
  \centering
  \leavevmode\epsfysize=6cm \epsfbox{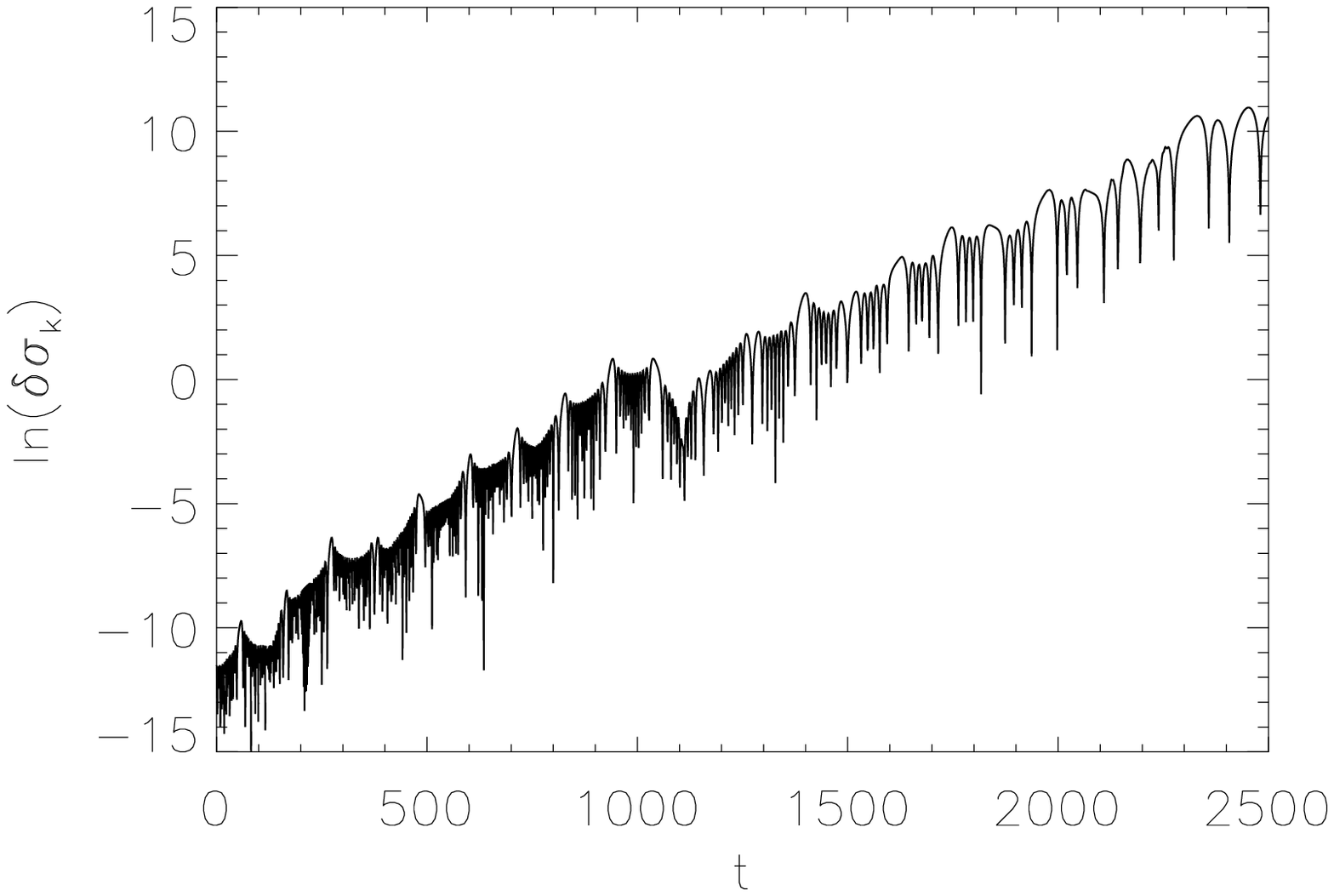}\\
\caption[fig1]{The evolution of $\ln \delta \sigma $ for 
  $g\frac{\phi _0 ^2}{m_{\phi }^2} =9.10^3$, $k=0$ and $\gamma =1.22$
  corresponds to $D=6$ extra compactified dimensions.  }
\end{figure}
\begin{figure}[t]
  \centering
  \leavevmode\epsfysize=6cm \epsfbox{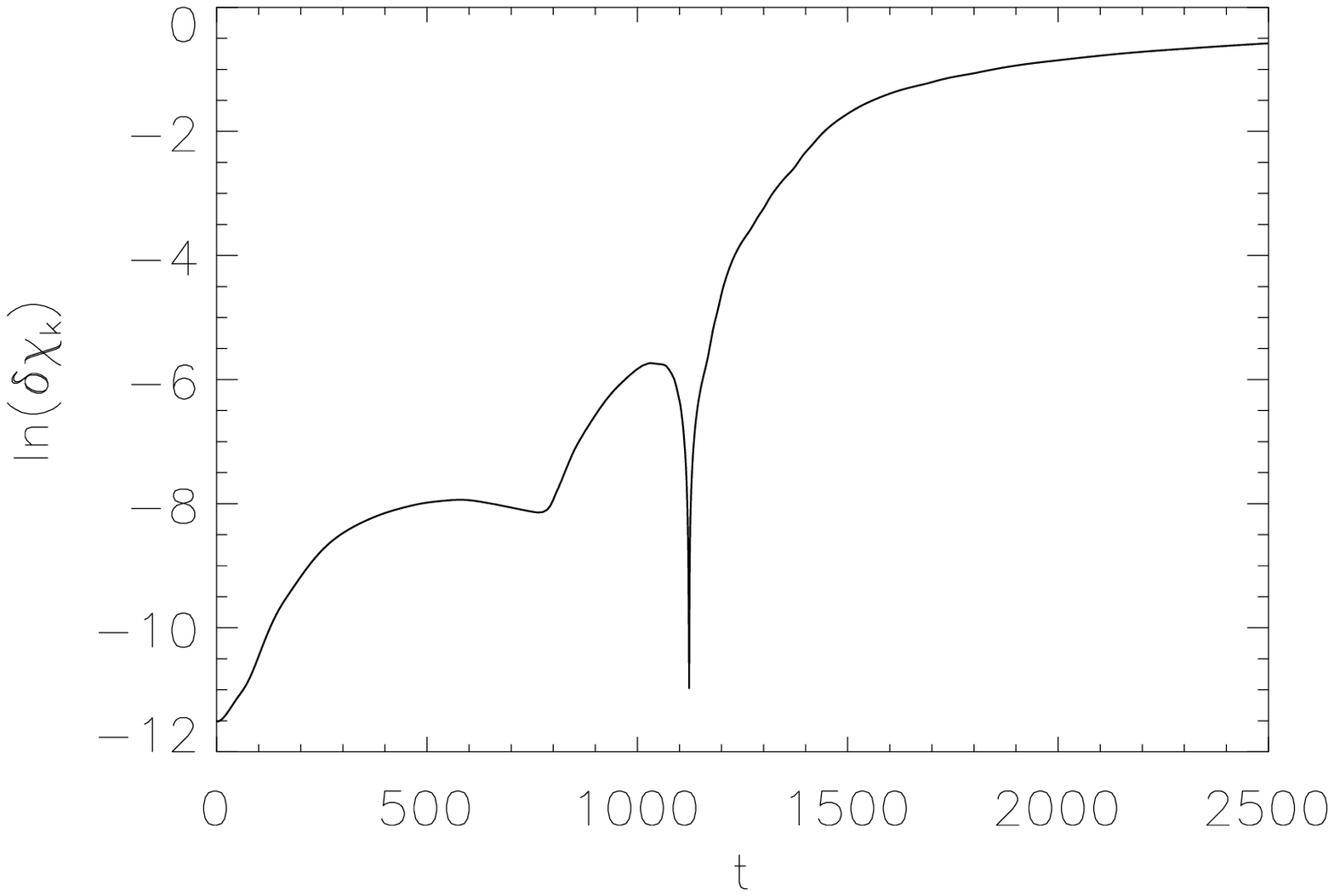}\\
\caption[fig1]{The evolution of $\ln \delta \chi $ for 
  $g\frac{\phi _0 ^2}{m_{\phi }^2} =9.10^3$, $k=0$ and $\gamma =0.09$
  corresponds to $\omega =500$ in JBD theory.  }
\end{figure}
\begin{figure}[t]
  \centering
  \leavevmode\epsfysize=6cm \epsfbox{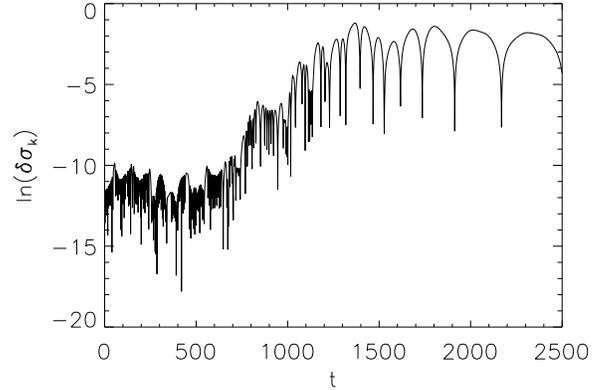}\\
\caption[fig1]{The evolution of $\ln \delta \sigma $ for
  $g\frac{\phi _0 ^2}{m_{\phi }^2} =9.10^3$, $k=0$ and $\gamma =0.09$
  corresponds to $\omega =500$ in JBD theory.  }
\end{figure}

\subsection{Perturbing Newton's Constant}
In~\cite{stein} the authors have considered the oscillations in the
Newton's constant by introducing a potential for the BD field
explicitly and concluded that $G$ oscillates about its mean value and
for reasonable values of the mass of the BD field ($m \geq 1$ GeV),
the oscillations have very high frequency, $\nu^{-1} \ll 1$ s,
compared to the Hubble expansion. In their case however the amplitude
of the oscillations is exponentially small, and the oscillation energy
is dissipated through Hubble redshift. Considering our scenario,
where we do not have to invoke the potential for the BD explicitly,
the perturbations in the BD field will cause the oscillations in the
Newton's constant in the Jordan frame even if the modes are well
within the horizon. Such oscillations will not only have a temporal
but also a spatial variation. As we approach to $k=0$ mode, gradually
the amplitude of such oscillations also increases exponentially in
time and causes a large variation in the Newton's constant. Such
oscillations are certainly permissible but whether they would
withstand all known tests from cosmology and general relativity is not
known at the moment. It is expected that the classical fluctuations
will lead to non linearity after some time and the linear theory of
perturbation will break down when $\Phi =1$. Such provocative
interpretation of the Newton's constant and the non-decayable property
of the BD field could be the source of dark matter in our universe.

Had we introduced a potential with a minimum
for the BD/dilaton field, superimposed on the
exponential growth, we should also see small oscillations with
frequency and amplitude depending on the exact form of the potential. 
\subsection{Entropy Perturbations}
During inflation the presence of BD field renders the three-curvature of
comoving hypersurfaces in terms of Bardeen's gauge invariant quantity
$\zeta $ time varying on super horizon scales, thus producing not only
adiabatic but also isocurvature perturbations~\cite{starobinsky,garcia}. 
It has been estimated in~\cite{starobinsky} that at the end of 
inflation when $\chi $ is very close to zero the metric perturbation 
is dominated by the adiabatic contribution and at the time of horizon 
crossing the fractional energy density fluctuations in $\chi$ are
much smaller than the adiabatic ones. This suggests that the isocurvature
fluctuations in the JBD model are negligible. If we assume that the BD
field is a candidate for the cold dark matter then one can estimate
its energy density from the exact solution of the BD field during the dust
era, which is a good approximation during the oscillatory phase when
the average pressure becomes zero and energy density falls as $1/t$.
Since 
\begin{equation}
\label{energy}
\rho_{\chi } = \frac{1}{2}\dot \chi ^2 \cong \gamma ^2 \rho \,,
\end{equation}
where $\rho $ is the post inflationary energy density of the universe, it
is obvious that the energy density of the BD
field is negligible when compared to the total. Usually for multiple
fields the difference between the relative perturbations are defined
as entropy perturbation $S_{1,2}$~\cite{kod} by:
\begin{eqnarray}
S_{1,2}=\frac{\delta \rho_1}{\rho_1 +p_1}-\frac{\delta \rho_2}{\rho_2
  +p_2} \,.
\end{eqnarray}
where $\rho_1$, $\rho_2$, $p_1$, $p_2$ are respectively the energy
densities and pressures.  For non-adiabatic initial conditions $\delta \chi
/\dot \chi \not= \delta \phi /\dot \phi \not= \delta \sigma /\dot
\sigma $, $S_{\chi , \phi}$, $S_{\chi , \sigma}$ and $S_{\sigma
  ,\phi}$ would not vanish, thus producing entropy perturbation inside
as well as outside the horizon. Thus for superhorizon modes
isocurvature fluctuations would evolve though adiabatic fluctuation
would have frozen. The growth in the isocurvature perturbation outside
the horizon gives the tilt in the spectrum, which is usually blue
shifted and roughly estimated by the ratio of effective mass square of
the CDM field and square of the Hubble expansion. If $\chi $ is
treated as a CDM field then the effective mass is roughly $\gamma ^2
V(\phi ,\sigma)$, the main contribution coming from the large coupling
$0.5g\phi ^2 \sigma ^2$, but the coupling is subdued by the presence
of $\gamma ^2$, provided we are in a weak limit JBD theory. In string
motivated and in Kaluza-Klein theories the effective mass increases a
lot. During preheating the Hubble expansion is very small compared to
the effective mass and thus results in an extreme blue tilt specially
in string motivated theories.
\subsection{Shaking the Kaluza-Klein Modes}
As we have described, depending on the numerical value of the coupling
constant, $\gamma $, the role of $\chi $ also changes, and in fact it covers a 
wide range of theories from effective action for superstring models to JBD
theories. If $\Phi _{BD}$ is interpreted as a homogeneous field which appears
as an effective mass term for the Kaluza-Klein modes \cite{shinji} in the 
effective four dimensional Kaluza-Klein action defined as
$M^2_{l}(\Phi_{BD})$ then, as mentioned before, the BD/dilaton can be 
related to the 
radius of the compactified $D$ dimensions if the compactification is done on a 
$D$ dimensional torus or on sphere. In terms of $\chi $: 
\begin{eqnarray}
\chi &=& \chi_{0} \ln\left[\frac{b}{b_0}\right] \,, \\
\chi_0 &=& \sqrt{\frac{D(D+2)}{2}}\,.
\end{eqnarray}
If the extra dimensions are compactified on a sphere then the 
mass $M_{l}(\chi )$ of the four dimensional Kaluza-Klein field $\phi_{lm}$
in terms of $\chi $ 
\begin{equation}
\label{Kalu}
M^2_{l}(\chi ) = \frac{l(l+D-1)}{b^2_0}e^{-(D+2)\chi /\chi_0 } \,.
\end{equation}
In the usual case, $\chi $ is assumed to be very close to zero after inflation 
and its variation is assumed to be negligible in radiation and matter dominated
eras. That leads to very high mass of these modes, very close to Planck scale
and proportional to the inverse square of the radius of the extra 
dimensions.
During reheating the perturbation in $\chi $ grows exponentially in
time and that
causes fluctuations in the mass of the Kaluza-Klein modes. The perturbations
effectively give rise to the Casimir force as a fluctuating boundary condition
for the $D$ dimensional sphere. This in fact can stabilize the
potential for the homogeneous scalar 
field, in our notation $\Phi_{BD}$. Though we have neglected such potentials in
our analysis since we have concentrated on the compactification on a 
torus, in this case the Kaluza Klein mass scale is still inversely 
proportional to the square of the size of the compactification radius.
If the compactification is done on a sphere and if a potential term for
the dilaton is 
taken into account then the effect of resonance will be even more
pronounced. However, the potential lacks a local minima and
in order to stabilize it one needs to invoke the Casimir effects
at a $1$-loop level. The issue raises a few questions such as particle creation
in the external dimensions due to fluctuating boundary and their stability 
requires a detailed study. 

If the Kaluza-Klein modes are excited their excitations may also give rise to 
some undesirable particles such as the pyrgon \cite{slan}, whose
energy density 
decreases as $a^{-3}$, giving during the radiation era $\rho_{\rm
pyrgon}/\rho_{\rm rad} \sim a$. Their 
stringent requirement to decay~\cite{slan} 
can enhance their life time and they can be
a candidate for dark matter. Overproduction of such particles could have 
overcome the critical density, but their number density could have been 
diluted by invoking a secondary phase of inflation. Such a phase is
also desirable 
to smoothen the large super horizon resonances as mentioned in \cite{bassett1}.
\section{Conclusions}
\label{sec:conc}
We have studied the preheating scenario in the context of
scalar-tensor theories. We have discussed the linear perturbation
theory and showed that the perturbations in BD/dilaton field grow
outside the horizon. The interpretation of such modes is not very
clear at the moment but one may hope that by including backreaction in
a consistent way may solve this problem. Nevertheless, including 
perturbations in the BD/dilaton field introduces new physics well
within the horizon such as temporal and spatial variation in Newton's 
constant. Such perturbations could play an important role
during structure formation and it is important to see whether such
variations in $G$ could be detected or not. In the strong coupling
regime fluctuations in $\chi$ field grow faster than the fluctuations
in the case of weak coupling (JBD theory) for the zero mode and can
also excite the Kaluza-Klein modes by lowering the
mass of $\phi_{l,m}$. Such excitations can give rise to highly
non-interacting quanta which can be a very good candidate for the
present dark matter. 

We have also discussed the entropy perturbations
during reheating which are solely due to the fact that there is more
than one scalar field and the initial conditions in the relative
density fluctuations in the respective components are non-adiabatic.
In general relativity the reason behind such super amplification in
the fluctuations is due to the enhancement in the entropy
perturbations outside the horizon. In our case such amplification is
even stronger especially in the strong coupling case because we have
an extra field which contributes its fluctuations to the entropy
perturbation. Isocurvature fluctuations will be
generated with a large tilt in string motivated and Kaluza-Klein
theories provided BD/dilaton is treated as a CDM field.

Here we must point out that we have intentionally neglected the potential 
term coming from the dilaton sector. Such a scheme is possible provided the
extra dimensions are compactified on a torus and not on a sphere.
Compactifying on a sphere gives rise to a term proportional to the
curvature of the sphere and such a term acts as a potential for the
dilaton. However, as we noted before, the curvature term lacks the
global minima in the direction of the dilaton
field and one has to invoke the first order Casimir corrections at a
1-loop level to give rise to a global minimum. Parametric excitations of
the Kaluza-Klein modes in such a potential have already been discussed
in~\cite{shinji} but the author has not taken the metric perturbations
into account. Inclusion of the dilaton potential may even enhance the
metric perturbation as it acts as an extra source term for 
Eq.~(\ref{pert4}). Concrete predictions requires a detailed study and it
is worth investigating this point as a separate issue. We must
mention that recently there has been an intense discussion upon the
initial conditions for the perturbations in the matter and in the
BD/dilaton fields. We have chosen our initial conditions to be scale
invariant. There have been other proposals such as the commonly used
quantum plane wave initial conditions for the fluctuations in the
fields. This however favours small $k$ modes more than the large $k$
modes. Another attractive proposal can be evolving the BD/dilaton from
$60$ e-foldings before the end of inflation when $aH \ll k$ until the
horizon crossing and then subsequently evolving the fields during the
oscillatory phase. These are open issues which require further 
investigation. Our analysis was mainly restricted to the
$k\rightarrow 0$ mode but we should also expect the same physical
consequences for other modes excited during preheating although with a
much smaller magnitude.

The issue of backreaction becomes important after a few inflaton 
oscillations and the subject becomes almost intractable as we increase 
the number of fields gradually. Backreaction has been considered 
before by many authors \cite{linde}, but in our situation we have to
worry about the contribution coming from the BD/dilaton sector also. A
self consistent study of the inflaton and the $\sigma$ particles should
take into account the effective change in the mass of the inflaton 
but in this case such corrections will be larger due to the
contribution from the BD/dilaton sector. Similarly, 
the BD/dilaton will also acquire an effective mass. To make a complete
picture one must answer all these relevant issues and we leave them 
for future investigation.
We also point it
out that our analysis has ignored the 
the contribution coming from $\Phi_{k=0}$ to the background evolution
equations. Inclusion of such terms will certainly help to amplify the 
fluctuations even faster and we can expect the fluctuations to become
non-linear before. The aspects of non-linearity in the gravitational sector
is least understood and the further evolution of the non-linear modes 
will become important leaving an imprint on the cosmic microwave background
radiation which will be the ultimate verification for the validity of
such claims.

\section*{Acknowledgments}

A.M. is supported by the Inlaks foundation and the ORS award. L.E.M.
is supported by FCT (Portugal) under contract PRAXIS XXI BPD/14163/97.
We are grateful to Andrew Liddle for discussions on various aspects of
perturbation theory. We also thank Juan Garc\'{\i}a-Bellido and Ian
Grivell for stimulating discussions. L.E.M acknowledges discussions
with Alfredo Henriques and Gordon Moorhouse.

\end{document}